\documentclass[prb,twocolumn,preprintnumbers,amsmath,amssymb,nopacs,nofootinbib,floatfix,superscriptaddress]{revtex4}

\usepackage{graphicx,bm,xcolor}

\makeatletter
\def\graphicscale{\twocolumn@sw{0.3}{0.4}}
\def\graphicthreescale{\twocolumn@sw{0.3}{0.4}}

\begin{document}

\title{Out-of-equilibrium dynamics driven by localized time-dependent 
perturbations \\ at quantum phase transitions }

\author{Andrea Pelissetto}
\affiliation{Dipartimento di Fisica dell'Universit\`a di Roma ``La Sapienza" 
  and INFN, Sezione di Roma I, I-00185 Roma, Italy}

\author{Davide Rossini}
\affiliation{Dipartimento di Fisica dell'Universit\`a di Pisa
        and INFN, Largo Pontecorvo 3, I-56127 Pisa, Italy}

\author{Ettore Vicari}

\affiliation{Dipartimento di Fisica dell'Universit\`a di Pisa
        and INFN, Largo Pontecorvo 3, I-56127 Pisa, Italy}

\date{\today}

\begin{abstract}
We investigate the quantum dynamics of many-body systems subject to
local, i.e., restricted to a limited space region, time-dependent
perturbations.  If the system crosses a quantum phase transition, an
off-equilibrium behavior is observed, even for a very slow driving.
We show that, close to the transition, time-dependent quantities obey
scaling laws. In first-order transitions, the scaling behavior is
universal, and some scaling functions can be exactly computed. For
continuous transitions, the scaling laws are controlled by the
standard critical exponents and by the renormalization-group dimension
of the perturbation at the transition.  Our protocol can be
implemented in existing relatively small quantum simulators, paving
the way to quantitatively probe the universal off-equilibrium scaling
behavior, without the need to manipulate systems close to the
thermodynamic limit.
\end{abstract}

\pacs{05.30.Rt,64.70.qj,64.60.an}

\maketitle



\section{Introduction}
\label{intro}

Quantum phase transitions (QPTs) are one of the most striking
signatures of many-body collective behavior, tantalizing the attention
of a large body of theorists and experimentalists working in condensed
matter and statistical physics.\cite{Sachdev-book}  The standard
paradigm of a QPT foresees a drastic change of the structural
properties of the system at zero temperature, when a given parameter
in the Hamiltonian is tuned across some critical value.  Generally,
the driving parameters are global homogeneous quantities coupled to
the critical modes, such as the magnetic field in spin
systems,\cite{Pfeuty-70, Coldea-10, Simon-etal-11, Islam-etal-11} the
chemical potential in particle systems,\cite{Fisher-89, Greiner-02,
  Bakr-10} etc.  However, close to a first-order transition, where
equilibrium low-energy properties are particularly sensitive to
localized external fields and/or boundary conditions, QPTs may also be
driven by local perturbations.\cite{CPV-15}

It is also tempting to study the dynamics across QPTs, induced by
time-dependent parameters. Under these conditions, the system is
inevitably driven out of equilibrium, even when the time dependence is
very slow, because large-scale modes are unable to equilibrate as the
system changes phase.  Off-equilibrium phenomena, as for example
hysteresis and coarsening, Kibble-Zurek defect production, aging,
etc., have been addressed in a variety of contexts, both
experimentally and theoretically (see, e.g., 
Refs.~\onlinecite{Binder-87,Bray-94,CG-04,BVS-06,Dziarmaga-10,PSSV-11,%
GAN-14,Biroli-15,Ulm-etal-13,LDSDF-13,NGSH-15} and references therein).  These
studies mostly focused on the effects of slow changes of global
parameters across classical and quantum transitions.  They have shown
that time-dependent properties of systems evolving under such dynamics
obey off-equilibrium scaling (OS) behaviors, depending on the
universal static and dynamic exponents of the equilibrium
transition.\cite{Polkovnikov-05, ZDZ-05, Dziarmaga-05, GZHF-10,
  CEGS-12, PV-17}

Here we overcome this paradigm and consider quantum systems subject to
a local, i.e., restricted to a limited space region, time-dependent
driving.  We investigate whether and how these perturbations bring the
system out of equilibrium as it moves across the different phases,
showing the emergence of a universal behavior, as observed in the case
of a global driving.  Our analysis provides a very intuitive and
simple framework enabling to develop a general OS theory that applies
both to first-order and continuous quantum transitions (FOQTs and
CQTs, respectively).  The beauty of this approach resides in the
possibility to quantitatively test universal quantum behavior even in
a relatively small setting,\cite{Zhang-17} without the need of
much larger sizes approaching the thermodynamic limit (as, e.g., for
the Kibble-Zurek framework), which would limit the experimental control
over the sample and prevent from a quantitative testing.
In view of the recent groundbreaking advancements in the field of quantum
simulation, these issues acquire specific relevance as a proposal
for experiments with a minimal set of requirements: manipulating and controlling
individual quantum objects, without the need of scalability.\cite{GAN-14}

To fix the ideas, we concentrate on the quantum Ising ring, a
paradigmatic model which undergoes various FOQTs and CQTs, when
varying its parameters.\cite{Sachdev-book} We present analytical and
numerical results for the off-equilibrium behaviors arising from slow
time-dependent protocols associated with local perturbations at its
quantum transitions. They support the general OS arguments developed
in the paper.

The paper is organized as follows.  In Sec.~\ref{isring} we introduce
the quantum Ising ring model, and review its equilibrium behavior in
the presence of local (constant) perturbations at the quantum
transitions. In Sec.~\ref{OFSSfo} we develop the OS theory for slow
time-dependent protocols associated with local perturbations at FOQTs;
the OS functions of the quantum Ising ring along the FOQT line are
computed by a two-level approximation, which turns out to be
asymptotically exact.  In Sec.~\ref{conttra} we extend our study
of the effects of
time-dependent local perturbations at the CQT of the quantum Ising
ring, showing that they give rise to OS behaviors as well.  In
Sec.~\ref{magkink} we study the off-equilibrium dynamics at the
magnet-to-kink transition arising when a local bond perturbation is
tuned along the FOQT line of the quantum Ising ring.  Finally,
Sec.~\ref{conclu} presents a summary and our conclusions. In the
appendix we focus on the dynamic two-level reduction exploited to
compute the OS functions of the quantum Ising ring along the FOQT
line.

\section{The quantum Ising ring}
\label{isring}

The quantum Ising Hamiltonian for a ring of $L$ sites is given by
\begin{eqnarray}
H = - \sum_{x=0}^{L-1} \left[ J \,\sigma^{(3)}_x \sigma^{(3)}_{x+1} 
+ g\, \sigma^{(1)}_x  
+ h \,\sigma^{(3)}_x\right]\, .
\label{hedef}
\end{eqnarray}
The spin-$1/2$ variables ${\bm \sigma}\equiv
(\sigma^{(1)},\sigma^{(2)},\sigma^{(3)})$ are the usual Pauli matrices,
and ${\bm \sigma}_{L} = {\bm \sigma}_{0}$.  We assume $\hslash=1$, $J=1$,
and $g>0$.  At $g=1$ and $h=0$ the model undergoes a CQT belonging to
the two-dimensional Ising universality class, separating a disordered
phase ($g>1$) from an ordered ($g<1$) one. For any $g<1$, the field
$h$ drives FOQTs, due to the crossing of the two lowest-energy states
$| + \rangle$ and $| - \rangle$ for $h=0$.  Correspondingly, the
longitudinal magnetization
\begin{equation}
M = {1\over L} \sum_{x=0}^{L-1} M_x,\qquad 
M_x\equiv \langle \sigma_x^{(3)} \rangle,
\label{longmag}
\end{equation}
is discontinuous, i.e.,\cite{Pfeuty-70}
\begin{equation}
\lim_{h\to 0^\pm} \lim_{L\to\infty} M = \pm m_0,
\qquad 
m_0 = (1 - g^2)^{1/8},
\label{m0def}
\end{equation}
and $\langle \pm | \sigma_x^{(3)} | \pm \rangle = \pm \,m_0$.  In a
finite system of size $L$, the lowest states are superpositions of $|
+ \rangle$ and $| - \rangle$, due to tunneling effects. For $h=0$,
their energy difference $\Delta$ vanishes exponentially as $L$
increases, $\Delta \sim g^L$, while the differences $\Delta_i\equiv
E_i-E_0$ for the higher excited states ($i > 1$) are finite for $L\to
\infty$. In particular, for the quantum Ising ring (corresponding to a
chain with periodic boundary conditions),\cite{Pfeuty-70,CJ-87}
\begin{eqnarray}
\Delta \equiv \Delta_1(L) \approx 2 \left({1-g^2\over \pi L}\right)^{1/2}
  g^L.
\label{deltalas}
\end{eqnarray}
The difference $\Delta_i$ for the higher excited states ($i > 1$)
remains finite for $L\to \infty$, at any value of $g \neq 1$. In particular
\begin{equation}
\Delta_{2}(L)=4(1-g)+O(L^{-2}).
\label{delta2}
\end{equation}
Conversely, for $g=1$, $\Delta_2(L) = \pi/(2L) + O(L^{-2})$.

In the following, we wish to analyze the quantum dynamics in the presence
of a single-site perturbation, adding
\begin{eqnarray}
H_s(t) = - s(t) \, \sigma_0^{(3)},\label{sitep}
\end{eqnarray}
to the Hamiltonian (\ref{hedef}) with $h=0$.  The {\em control
  parameter} $s(t)$ plays the role of a longitudinal magnetic field
acting on one site only.

Before discussing the effects of a time-dependent
perturbation, it is useful to summarize the equilibrium properties
of the model when $s(t)$ is constant, $s(t) = s$.  
In the disordered phase ($g<1$), the impact of the single-site
perturbation is expected to be limited, being restricted within a
region of finite size $\xi$.  Therefore, for large-scale bulk
quantities, the perturbation gives only rise to $O(\xi/L)$ corrections in
the large-$L$ limit.

Approaching the CQT, i.e., for $g\to 1^+$, the system develops
long-distance correlations, and $\xi$ diverges as $\xi \sim
(g-1)^{-\nu}$ with $\nu=1$.  Around $g=1$, the interplay between $\xi$
and $L$ originates an equilibrium finite-size scaling (EFSS) 
behavior.\cite{Privman-90,CPV-14} The effects of the local perturbation are
amplified by long-distance correlations.  Although they do not alter
the leading power-law behavior, scaling functions acquire a nontrivial
$s$-dependence.  Moreover, local quantities acquire a nontrivial
$x$-dependence.  For instance, the local magnetization $M_x$  is
expected to scale
as:\cite{Binder-DL,Barber-DL,Diehl-DL,CL-91,CZ-94,SM-12}
\begin{equation}
M_x(L,g\!=\!1,s) 
\approx L^{-\beta/\nu} {\cal M}_E(x_p/L,\xi/L,sL^{y_s}),
\label{mxlgscqt}
\end{equation}
where $x_p = \min(x,L-x)$ is the distance along the ring, $\beta=1/8$
is the magnetization exponent,\cite{Sachdev-book} $y_s=1/2$ is the
scaling dimension associated with the single-site parameter
$s$.\cite{CL-91,CZ-94} Thus, the average magnetization 
behaves as
\begin{equation}
M(L,g\!=\!1,s) \approx L^{-\beta/\nu} Q_E(sL^{y_s}).
\label{mlgqe}
\end{equation}

Along the FOQT line ($g<1$) the system is particularly sensitive to
local defects and boundary fields.\cite{CNPV-14,CPV-15} Indeed, the
single-site perturbation $H_s$ can control the bulk phase: as $s$
changes sign, the bulk magnetization $M$ switches from $-m_0$ to
$m_0$.  An EFSS behavior can be defined at FOQTs,\cite{CNPV-14}
analogously to CQTs.
In the case at hand, the relevant scaling variable is 
\begin{equation}
\kappa = {2 m_0 s\over \Delta}, 
\label{kappadef}
\end{equation}
where $\Delta$ is the gap for $s\!=\!0$, defined in Eq.~\eqref{deltalas},
so that 
\begin{equation}
M(L,g,s) \!\approx m_0 f_E(\kappa)
\label{mlgs}
\end{equation}
for any $g<1$.

The EFSS functions can be obtained by performing a
two-level truncation, keeping only the lowest levels $|\pm \rangle$.
This approximation holds whenever the energy difference between two
such states remains much smaller than those between the higher
excited states and the ground state.
This requires 
\begin{equation}
{\Delta\over \Delta_2} \approx {1\over 2}
 \left[ {1+g\over (1-g)\pi L}\right]^{1/2}
g^L  \ll 1 , \label{ratio}
\end{equation}
for $s=0$, and  
\begin{equation}
m_0 |s|\ll \Delta_2 \approx 4 ( 1 - g), \label{m0bd}
\end{equation}
where we used the asymptotic behaviors of $\Delta$ and $\Delta_2$ at
$s=0$, cf. Eqs.~(\ref{deltalas}) and (\ref{delta2}).
For generic
values of $g$, Eq.~(\ref{ratio}) is already satisfied for moderately
large sizes.  For example, for $g=1/2$, $\Delta/\Delta_2\approx
0.0068$ for $L=5$, and $\Delta/\Delta_2\approx 0.00067$ for $L=8$.

The Hamiltonian restricted to this subspace has the
form\cite{CNPV-14}
\begin{eqnarray}
H_e = \left(
\begin{array}{l@{\ \ }l@{\ \ }l}
\varepsilon  - \beta &  \quad \delta e^{i\varphi}  \\
\delta e^{-i\varphi}  &  \quad\varepsilon + \beta\\
\end{array} \right) \; , \label{hr}
\end{eqnarray}
where $\beta=m_0 s$ represents the perturbation induced by the local
magnetic field $s$, and $\delta=\Delta/2$ is a small parameter which
vanishes for $L\to \infty$ and $s=0$, giving rise to a degenerate
ground state.  The phase $\varphi$ is irrelevant, thus we can set
$\varphi=0$ (it can be absorbed in the definition of the states).  The
eigenstates of $H_e$ are $(0 < \alpha \le \pi/2$)
\begin{eqnarray}
&& |0\rangle = \sin(\alpha/2) \, |-\rangle +  
\cos(\alpha/2) \, |+\rangle, \label{eigstate0}\\
&& |1\rangle =  \cos(\alpha/2) |-\rangle -
\sin(\alpha/2) \, |+\rangle, \label{eigstate1}
\end{eqnarray}
where
\begin{equation}
\tan \alpha = \kappa^{-1}, \qquad \kappa = {\beta\over \delta} = {2m_0 s\over \Delta}.
\label{alphadef}
\end{equation}
Their energy difference is 
\begin{equation}
E_1 - E_0 =
\Delta \; \sqrt{1 + \kappa^2}.
\end{equation}
The magnetization is obtained by computing the expectation value of
$\sigma^{(3)}$ on the ground state $|0\rangle$,
\begin{eqnarray}
f_E(\kappa) = \cos \alpha = {\kappa\over \sqrt{1 + \kappa^2}}.
\label{fsigma}
\end{eqnarray}

In the following we discuss the quantum evolution of the Ising
model~\eqref{hedef} with $h=0$, in the presence of a local
longitudinal field~\eqref{sitep} obeying a linear time dependence
\begin{equation}
s(t) = c\, t,
\label{hst}
\end{equation}
with time scale $t_s \sim c^{-1}$.
The protocol starts at $t_i < 0$,
from the ground state at $s(t_i)=s_i<0$.  Then, the quantum dynamics
evolves up to $t=t_f>0$, $s(t_f)=s_f>0$, so that $s(t)$ crosses the
{\em critical} value $s=0$.
We compute observables, such as the magnetization and correlation
functions, during the quantum evolution both along the FOQT line
(Sec.~\ref{OFSSfo}), and at its endpoint $g=1, \, h=1$, where an
Ising CQT appears (Sec.~\ref{conttra}). 
We stress that our protocol (\ref{hst}) is quite general, since
arbitrary time dependences can be linearized around $s=0$. Below we 
comment more in depth on this point.

\section{Off-equilibrium finite-size scaling along the FOQT line}
\label{OFSSfo}

\subsection{Off-equilibrium finite-size scaling}
\label{proofss}

In this section we develop the off-equilibrium finite-size scaling
(OFSS) theory for the quantum evolution arising from the
time-dependent protocol associated with the local perturbation
(\ref{sitep}) along the FOQT line.  For this purpose we must identify
the relevant scaling variables. Since EFSS should be recovered in the
appropriate limit (defined below), one of them can be obtained from
the equilibrium variable $\kappa= 2 m_0 s/\Delta(L)$ by replacing $s$
with $s(t)=c\,t$,
\begin{equation}
\kappa \equiv {2 m_0 s(t)\over \Delta} = {2 t \over \Delta \, t_s},
\label{katdef}
\end{equation} 
where $t_s\equiv (m_0 c)^{-1}$.  A natural choice for a second OS
variable is 
\begin{equation}
\theta \equiv t\,\Delta.  
\label{thetadef}
\end{equation}
We also define the related OS variables
\begin{eqnarray}
&&\upsilon \equiv \Delta^2 t_s=2\theta/\kappa,
\label{upsdef}\\
&&\tau \equiv
{t/\sqrt{t_s}}={\rm sign}(\theta)\sqrt{\kappa\theta/2}.
\label{taudef}
\end{eqnarray}
The OS limit is defined by $t,t_s,L\,\to\infty$, keeping the above
OS variables fixed. In this limit, the magnetization is expected
to show the OFSS behavior
\begin{equation}
M(t,t_s,L) \approx m_0\,f_O(\upsilon, \kappa) = m_0\,F_O(\upsilon,\tau),
\label{mtsl}
\end{equation} 
where $\tau = \sqrt{\upsilon} \kappa/2$.  In the adiabatic limit
($t,t_s\to \infty$ at fixed $L$ and $t/t_s$) EFSS must be recovered,
so that 
\begin{equation}
f_O(\upsilon\to\infty, \kappa)=f_E(\kappa).
\label{folim}
\end{equation}

The OS behavior is expected to develop in a narrow range of $s(t)\approx 0$; indeed,
since $\tau$ is kept fixed in the OS limit and $s(t) \sim
\tau/\sqrt{t_s}$, the relevant interval of $s(t)$ decreases as $t_s$
increases. This implies that the OFSS behavior is independent of the
initial and final values of $s$.  The OS functions are universal,
i.e., independent of $g$ along the FOQT line. The approach to OFSS is
expected to be controlled by the ratio between $\Delta\sim e^{-cL}$
and $\Delta_2=O(1)$, cf. Eqs.~(\ref{deltalas}) and (\ref{delta2}),
therefore in the case of model~\eqref{hedef}, to be exponentially
fast.  We stress that the above arguments are quite general and can be
straightforwardly extended to any FOQT.

We may also consider a generic protocol characterized by the time
scale $t_s$, i.e., $s(t) =S(t/t_s)$ with $S(0)=0$ and $S'(0) \not=0$.
Since the OS limit is taken by keeping $\tau \equiv t/\sqrt{t_s}$
fixed, we can expand $S(t/t_s)$ in powers of $t/t_s = \tau/\sqrt{t_s}$
and only keep the leading term in the OS limit.  Higher-order terms
give $O(t_s^{-1/2})=O(\Delta)$ contributions: they are 
exponentially suppressed with the system size.

\subsection{Two-level approximation}
\label{twolev}

The OS functions at the FOQTs of the Ising ring can be exactly
computed.  Remarkably, in a way similar to EFSS,
in the long-time limit and for large systems,
the scaling properties in a small interval around $s=0$ (more
precisely, for $m_0 |s(t)|\ll \Delta_2$) are well captured by a
two-level truncation,\cite{CNPV-14} which only takes into account the
two nearly-degenerate lowest-energy states. This is demonstrated in
App.~\ref{twoldyn}.

The effective evolution is determined by the Schr\"odinger equation 
\begin{equation}
i \, \partial_t \Psi(t) = H_r(t) \Psi(t), 
\label{hrdef}
\end{equation}
where $\Psi(t)$ is a combination of the states $|+ \rangle$ and
$|-\rangle$ only.  The time-dependent Hamiltonian $H_r(t)$ can be
determined by assuming that its matrix elements are analytic functions
of $t/t_s$, and that the nonanalyticity only arises in the limit $L\to
\infty$, when the two states become degenerate.  Using the symmetry of
the model, one can see that it is enough to consider 
\begin{equation}
H_r = -{t\over
  t_s} \sigma^{(3)} + {\Delta\over 2} \sigma^{(1)},
\label{hrdef2}
\end{equation}
where $\Delta/2=\delta$ is the same amplitude that enters the
off-diagonal terms of Eq.~\eqref{hr}.

The dynamics is analogous to that governing a two-level quantum
mechanical system in which the energy separation of the two levels is
a function of time, which is known as the Landau-Zener (LZ)
problem.\cite{LZeff}  Therefore, the
time-dependent wave function for the quantum Ising ring can be derived from 
the LZ corresponding solutions.\cite{VG-96}  We obtain
\begin{equation}
\Psi(t) = C_-(\upsilon, \tau) |-\rangle + C_+(\upsilon,\tau) |+\rangle
\label{psisol}
\end{equation}
where $C_\pm$ are functions of the scaling variables $\upsilon = t_s
\Delta^2$ and $\tau ={t/\sqrt{t_s}}$.  In
particular, assuming $\Psi(\tau_i) = |-\rangle$ for $\tau_i=-\infty$, 
the OFSS function $F_O$ of the magnetization defined in Eq.~(\ref{mtsl}) 
is given by
\begin{eqnarray}
F_O(\upsilon,\tau)  &=& \langle \Psi(t) |\sigma^{(3)}|\Psi(t)\rangle
\label{fsigmasol}\\
&=& |C_+(\upsilon,\tau)|^2 - 
|C_-(\upsilon,\tau)|^2  \nonumber \\
&=& {\upsilon\over 4} 
e^{-{\pi \upsilon\over 16}} |D_{-1+i{\upsilon\over 8}}(\sqrt{2} 
e^{i{3\pi\over 4}}\tau)|^2 - 1,
\nonumber
\end{eqnarray}
where $D_\nu(z)$ is the parabolic cylinder function.\cite{Abrafunc}
By replacing $\tau$ with $\sqrt{\upsilon} \kappa /2$ in
Eq.~(\ref{fsigmasol}), we obtain the OS function
$f_O(\upsilon,\kappa)$ defined in Eq.~(\ref{mtsl}):
\begin{equation}
f_O(\upsilon,\kappa)=F_O(\upsilon,\sqrt{\upsilon}\kappa/2).
\label{foFo}
\end{equation}

Note that the initial condition $\Psi(\tau_i=-\infty)=|-\rangle$ is
consistent with the choice of the initial condition for the
time-dependent protocol (\ref{sitep}), i.e.,  $\Psi(s_i<0) =
|-\rangle$, because, when both $L$ and $t_s$ are large, any finite
$s_i<0$ is in the adiabatic region, where the ground state is given by
$\Psi(s_i)\approx |-\rangle$ with exponential precision. Indeed, a
finite $s_i$ corresponds to $\kappa\to -\infty$ in the large-$L$ and
$t_s$ limit keeping $\Delta^2 t_s$ finite, and for $\kappa\to -\infty$
the ground state (\ref{eigstate0}) is just given by $|-\rangle$.

\begin{figure}
\includegraphics[width=7.5cm]{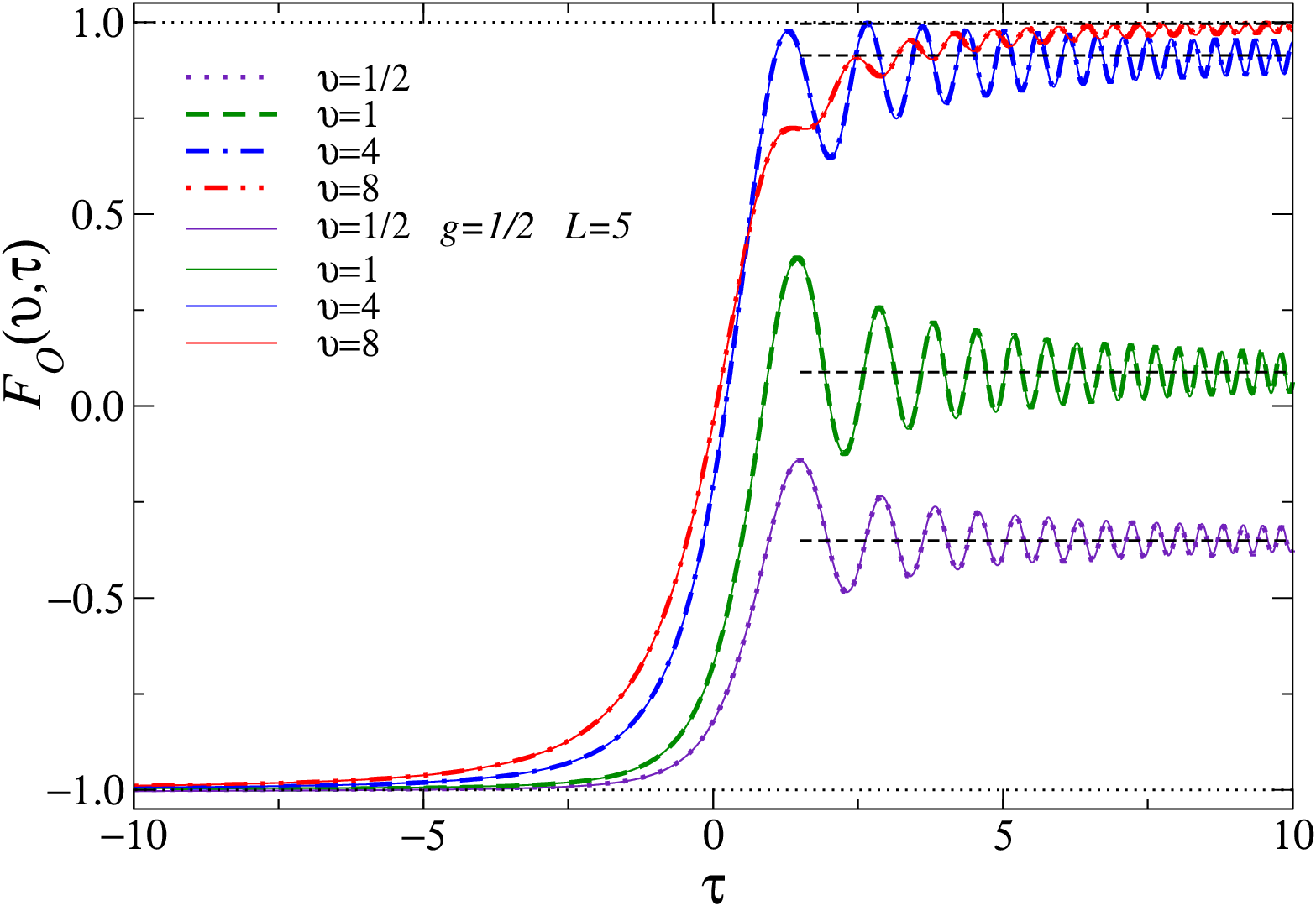}
\caption{The magnetization scaling function $F_O(\upsilon,\tau)$
  reported in Eq.~(\ref{fsigmasol}), for several values of $\upsilon$.  We
  also plot the corresponding time evolution of the ratio $M/m_0$ for
  the Ising ring at $g=1/2$ and $L=5$, under the protocol
  (\ref{hst}). Differences are hardly visible.  Note the oscillating
  behavior for $\tau>0$ around the asymptotic large-$\tau$
  value~\cite{LZeff} (short horizontal dash lines),
  $F_O(\upsilon,\tau\to +\infty) = 1 - 2 \,e^{-{\pi \upsilon/4}}$; the
  amplitude of the oscillations slowly decreases with increasing
  $\tau$.}
\label{fsigmafig}
\end{figure}

Plots of the function $F_O(\upsilon,\tau)$ for some values of
$\upsilon$ are shown in Fig.~\ref{fsigmafig}.  We have also
numerically computed the magnetization $M(t,t_s,L)$ for the quantum
Ising ring:\cite{footnotenum} the results displayed in
Fig.~\ref{fsigmafig} are in remarkable agreement with
Eq.~(\ref{fsigmasol}), even for small system sizes (we report data for
$L=5$), reflecting the exponentially fast convergence to the
asymptotic behavior. This validates the analytic derivation based on
the two-level approximation.

Several notable limits of the solution (\ref{fsigmasol}) can be
derived using the known properties of the parabolic cylinder 
function\cite{Abrafunc,VG-96}
$D_\nu(z)$. Concerning the $\tau$-dependence at fixed $\upsilon$, we obtain
\begin{eqnarray}
&& F_O(\upsilon, \tau\to - \infty) = - 1,\label{fsmin}\\
&& F_O(\upsilon,\tau=0)
= - \,e^{-{\pi \upsilon/8}}, \label{fst0}\\
&& 
F_O(\upsilon,\tau\to +\infty)
= 1 - 2 \,e^{-{\pi \upsilon/4}}. 
\label{lzlimit}
\end{eqnarray}
Note that the large-$\tau$ behavior is obtained by using the well known
KZ result 
for the transition probability from the ground state $|0\rangle$ to
the excited state $|1\rangle$, which is given by $|C_-(\upsilon,\tau\to
+\infty)|^2=e^{-\pi \upsilon/4}$.  The large-$\upsilon$ asymptotic
behavior of $F_O(\upsilon,\tau)$ at fixed $\tau$ is 
\begin{equation}
F_O(\upsilon\gg 1,\tau)\approx {2 \tau\over \sqrt{\upsilon + 4 \tau^2}},
\label{asyulfs}
\end{equation}
so that $F_O(\upsilon,\tau)$ trivially vanishes for $\upsilon \to \infty$
and any finite
$\tau$.  The limit $\upsilon\to 0$ corresponds to the infinite-volume
limit. We find
\begin{equation}
F_O(\upsilon\to 0,\tau) = -1.
\label{asyusfs}
\end{equation}
This reflects the effective decoupling of the states $|\pm\rangle$ of
the Hamiltonian (\ref{hrdef2}), which evolve independently in the
infinite-volume limit.

\begin{figure}
\includegraphics[width=7.5cm]{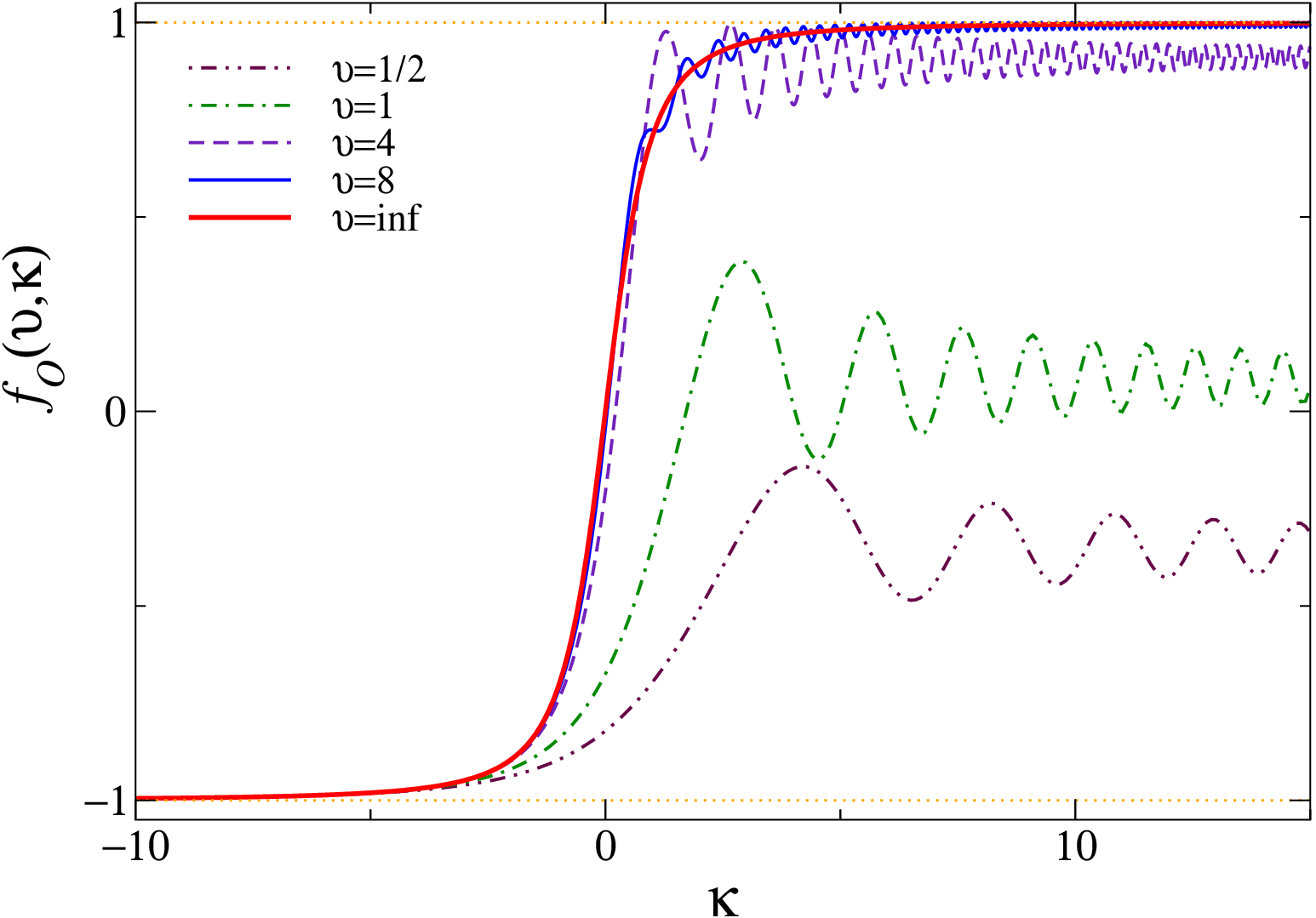}
\caption{ The scaling function $f_O(\upsilon,\kappa)$ associated with
  the magnetization, cf. Eq.~(\ref{foFo}), for some values of
  $\upsilon$.  In agreement with the OS arguments, it approaches the
  static limit $f_E(\kappa)=\kappa/\sqrt{1 + \kappa^2}$
  (red curve) when $\upsilon\to\infty$ keeping $\kappa$ fixed.}
\label{gsigmafig}
\end{figure}

In Fig.~\ref{gsigmafig} we show some plots of the function
$f_O(\upsilon,\kappa)$, obtained using Eq.~(\ref{foFo}). In agreement with the
OS arguments, it approaches the static limit when $\upsilon\to\infty$
keeping $\kappa$ fixed.  This is indeed confirmed by the solution
(\ref{foFo}). Replacing $\tau$ with $\sqrt{\upsilon}\kappa/2$ in 
Eq.~\eqref{asyulfs}, we find
\begin{equation}
f_O(\upsilon\to\infty,\kappa)= f_E(\kappa) = {\kappa\over 
\sqrt{1+\kappa^2}},
\label{gsigmaasyu}
\end{equation}
where $f_E(\kappa)$ is the EFSS function reported in Eq.~\eqref{fsigma}.

The fluctuations of the magnetization can be characterized by its
variance, given by
\begin{eqnarray}
V_M(\upsilon,\tau) &=& \langle \Psi(t) | (\sigma^{(3)})^2 | \Psi(t) \rangle -
\langle \Psi(t) |\sigma^{(3)}| \Psi(t) \rangle^2
\nonumber\\
&=& 1 - F_O(\upsilon,\tau)^2 \le 1.
\label{variance}
\end{eqnarray}
Note that $V_M$ is generally of the same size of $F_O$.

These results show that the OFSS functions at the FOQTs of the Ising
ring are well reproduced by the quantum dynamics of a two-level model,
although the OS variables $\kappa,\,\theta$ are determined by the
underlying many-body physics of the original model, which gives rise
to the exponential dependence of the gap $\Delta(L)$.  We stress that
these conditions are only realized when the many-body system is tuned
to the FOQT arising from a two-level crossing.  An analogous behavior
is expected in higher-dimensional quantum Ising systems.  

Since the
OFSS arguments leading to Eq.~\eqref{mtsl} apply to quite general
FOQTs, our protocol can be seen as a viable proposal for a controlled
quantum switch between the corresponding two states $|+\rangle$ and
$|-\rangle$ in the symmetry-broken phase of a few-spin Ising-like chain,
constituting an effective qubit.
This would enhance its robustness with respect to other local codings
(e.g., through the spin-degree of freedom of a single atom or molecule).
Switching from one to another state can be achieved by tuning a local
longitudinal
field, whose dynamical effects can be quantitatively controlled
by universal scaling functions.

We finally mention that an analogous behavior is expeted to emerge when
the external magnetic field is spatially uniform, i.e.  when
one adds the magnetic term
\begin{equation}
H_h = - h(t) \sum_{x=1}^L \sigma_x^{(3)},\qquad h(t) = a\, t,
\label{unht}
\end{equation}
instead of the local term (\ref{sitep}).  The protocol starts
at $t_i<0$ from the ground state at $h_i = h(t_i) < 0$,
which is again given by $|-\rangle$ in the
large-$L$ limit. Then, the system evolves up to a time $t_f > 0$
corresponding to a finite $h_f>0$.  The OS arguments apply here as
well. One should only change the definition of $\kappa$, considering 
\begin{equation}
\kappa = {2 m_0 h(t)L\over \Delta} \equiv {2 L t\over \Delta t_s},
\label{katdefunh}
\end{equation}
where we used the fact that the energy associated with the magnetic
perturbation is $E_h=m_0 h L$.  The second scaling variable is again
$\theta=\Delta t$, so that $\upsilon = (\Delta^2/L) t_s$. Using the
fact that $\Delta = a_g L^{-1/2} e^{-c_g L}$, we may write the scaling
variable corresponding to $\tau$, cf. Eq.~(\ref{taudef}), as $\tau
\approx t \;t_s^{-1/2} \ln t_s$.  Considerations based on the
effective two-level model lead to the OFSS behavior (\ref{mtsl}) for
the magnetization, with the same OS function $f_O(\upsilon,\kappa)$.

\section{Off-equilibrium dynamics at the continuous transition}
\label{conttra}

\begin{figure}
\includegraphics[width=7.5cm]{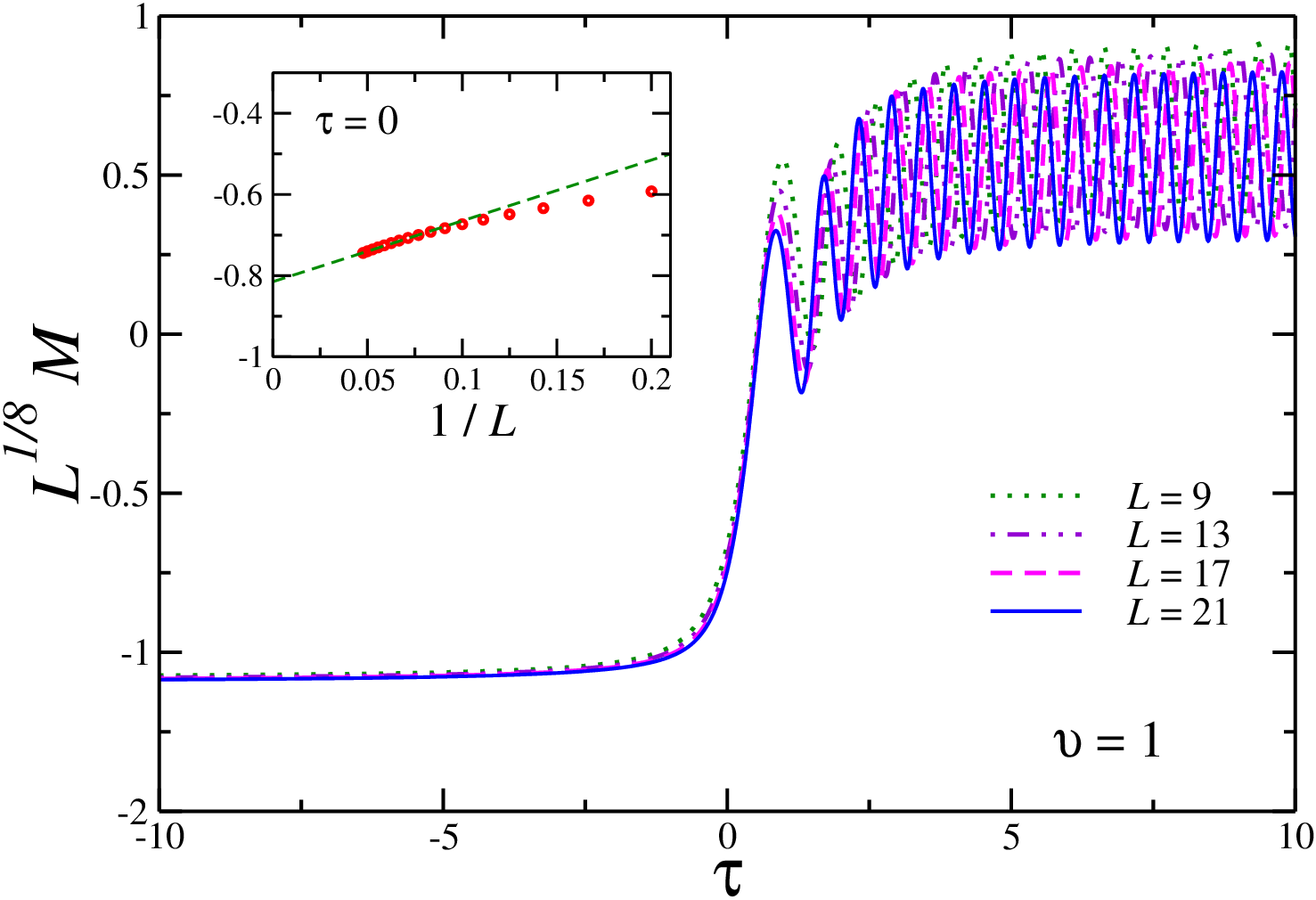}
\caption{Time-dependent magnetization at the CQT, in the presence of
  the single-site perturbation.  We plot $L^{1/8} M$ versus
  $\tau=t/t_s^{2/3}$ for $\upsilon\equiv t_s/L^{3/2}=1$ and several
  values of $L$, as obtained by numerical
  calculations.\cite{footnotenum}  The curves clearly approach an
  asymptotic function with increasing $L$, confirming the existence of
  OFSS. In the inset we show $L^{1/8} M$ for $t = \tau = 0$, as a
  function of $1/L$. }
\label{mscalingg1}
\end{figure}

It is interesting to compare the behavior along the FOQT line with
that occurring at its endpoint $g=1,\,h=0$, where a standard Ising CQT
occurs.  EFSS, Eq.~(\ref{mxlgscqt}), can be extended to the dynamic
case using scaling arguments analogous to those used at the FOQT.  The
scaling variables are $\kappa = (t/t_s) L^{y_s}$, with $y_s=1/2$
---this is the equilibrium scaling variable $s L^{y_s}$, in which we
have simply replaced $s$ with $s(t)$--- and $\theta=t \Delta\sim
t/L^z$ with $z=1$.  We also define the related OS variables 
\begin{equation}
\upsilon
\equiv t_s/L^{y_s+z},\qquad \tau \equiv {t/t_s^{z/(z+y_s)}}.
\label{conttrasca}
\end{equation}
Then, the local magnetization is expected to satisfy the OFSS equation
\begin{equation}
M_x(L,t,t_s)  \approx L^{-\beta/\nu} {\cal M}_O(x_p/L,\upsilon,\tau),
\label{mltts}
\end{equation}
so that its spatial average satisfies
\begin{equation}
M(L,t,t_s) \approx L^{-\beta/\nu} Q_O(\upsilon,\tau).
\label{mlttscon}
\end{equation}

These OS behaviors are confirmed by numerics on moderately large
systems,\cite{footnotenum} as displayed in Fig.~\ref{mscalingg1} for
$\upsilon = 1$ (analogous results are obtained for other values of
$\upsilon$). The inset shows that corrections decay as $1/L$.

\section{Off-equilibrium dynamics at the magnet-to-kink transitions}
\label{magkink}

Other interesting examples of QPTs driven by a local perturbation
arise when adding
\begin{eqnarray}
H_b(t) = b(t) \, \sigma_{\ell}^{(3)} \sigma_{\ell+1}^{(3)},
\qquad
\ell = \lfloor (L-1)/2 \rfloor,
\label{bondp}
\end{eqnarray}
to Hamiltonian \eqref{hedef} with $h=0$. In the static case, $b(t) =
b$, such term gives rise to CQTs when $g<1$, between two different
quantum phases:\cite{CPV-15} a {\em magnet} phase for $b<2$ and a
{\em kink} phase for $b > 2$.

In the magnet phase, the lowest states are superpositions of states
with opposite magnetization $|\pm\rangle $ (neglecting local effects
at the defect), and the gap is exponentially
small.\cite{ZJ-86,BC-87,CPV-15} In particular,~\cite{CPV-15}
\begin{equation} \Delta \approx {8g\over 1-g} w^2
e^{-wL}, \qquad w={1-g\over g} \;(2-b),
\label{defde}
\end{equation}
for $b \to 2^-$.  The large-$L$ two-point function, 
\begin{equation}
G(x_1,x_2) \equiv
\langle \sigma_{x_1}^{(3)} \sigma_{x_2}^{(3)}\rangle
\label{gdef}
\end{equation}
is trivially constant, i.e.,
\begin{equation}
G_r(x_1,x_2) \equiv {G(x_1,x_2)\over m_0^2} \to 1
\label{gcmagnet}
\end{equation}
for $x_1\neq x_2$, keeping $X_i\equiv x_i/\ell$ fixed.

The behavior drastically changes when $b > 2$, where the low-energy
states are one-kink states, which behave as one-particle states with
$O(L^{-1})$ momenta.  The ground state and the first excited state are
superpositions with definite parity of the lowest kink
$|\hspace{-0.5mm}\downarrow\uparrow\rangle$ and antikink
$|\hspace{-0.5mm}\uparrow\downarrow\rangle$
states.  The gap behaves as\cite{CPV-15}
\begin{eqnarray}
\Delta = {8 (b-1) g^2 \over b (b-2)(1-g)^2}\,{\pi^2\over L^3} +
O(L^{-4}).
\label{deltallm3}
\end{eqnarray}
Moreover, the two-point function $G(x,y)$ behaves asymptotically as
\begin{eqnarray}
{G(x_1,x_2)\over m_0^2} = 1 \!- \!|X_1-X_2| \!-\! {|\sin(\pi X_1)\!-\!\sin(\pi
  X_2)| \over \pi},
\label{gczetamm1}
\end{eqnarray}
where $X_i\equiv x_i/\ell$.

The parameter $b$ turns out to drive a CQT at $b = b_c =2$, separating
the magnet and kink phases,\cite{CPV-15} where the relevant scaling
variable is
\begin{equation}
\varepsilon_s \equiv \varepsilon\,L^{y_\varepsilon},\qquad
\varepsilon \equiv b-2,
\label{vareps}
\end{equation}
with $y_\varepsilon=1$. In the scaling limit, the two-point function
behaves as 
\begin{equation}
G(x_1,x_2) \approx m_0^2 \, {\cal
  G}(X_1,X_2;\varepsilon_s), 
\label{gscal}
\end{equation}
which implies 
\begin{equation}
\chi \equiv \sum_x G(0,x) = m_0^2 \,L\, f_\chi(\varepsilon_s).
\label{chisca}
\end{equation}
Since $\Delta \sim L^{-2}$ at $b_c$, this CQT has $z=2$ as dynamic
exponent.\cite{CPV-15}

We should emphasize that this transition is driven by a local
perturbation, contrary to the standard QPT paradigm, which requires a
global tuning.\cite{Sachdev-book}  The key point is again associated
with the underlying FOQT, which makes the system particularly
sensitive to local defects.

\begin{figure}
\includegraphics[width=7.5cm]{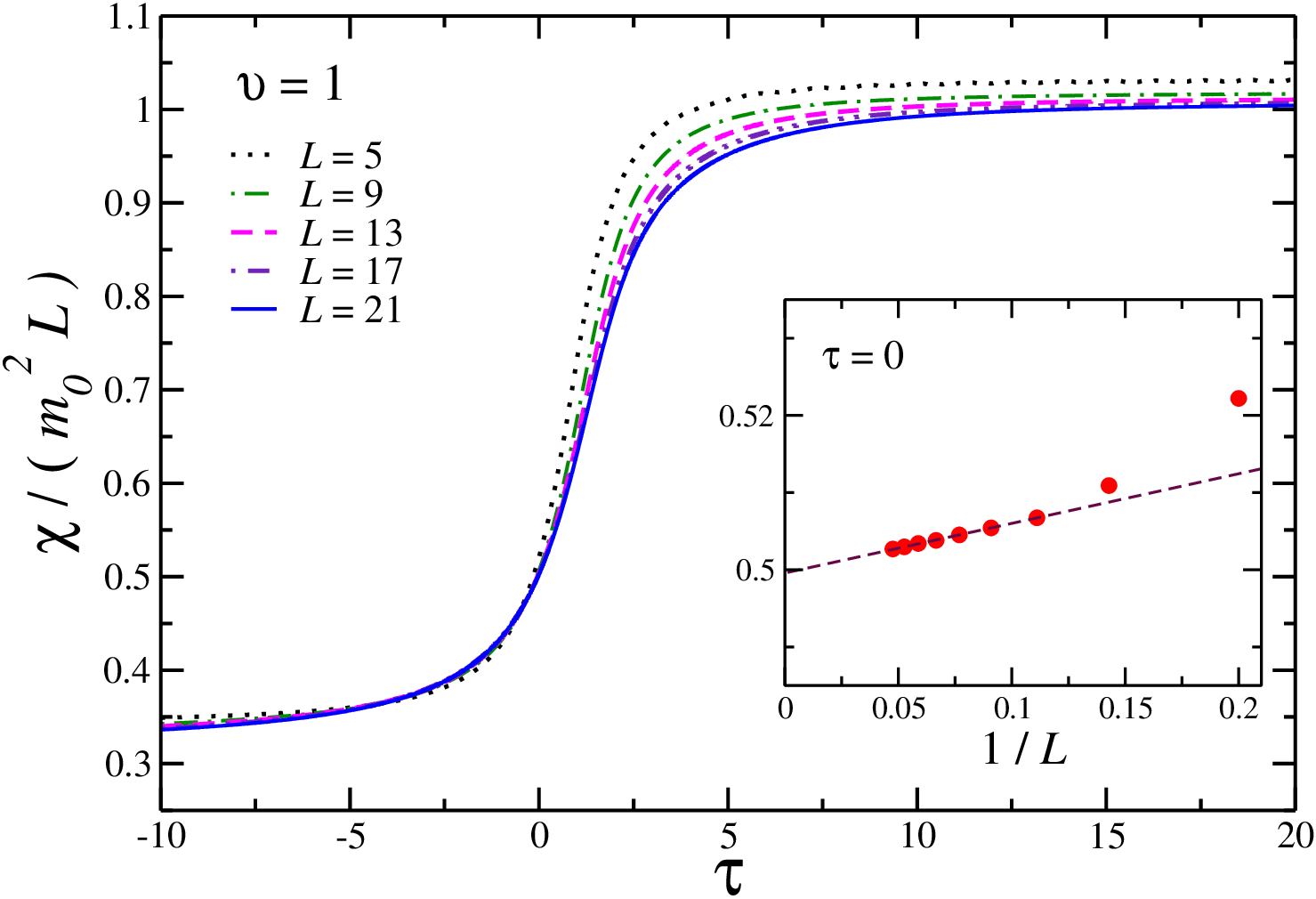}
\caption{Plot of $\chi/(m_0^2 L)$ versus $\tau=t/t_s^{2/3}$ for
  $\upsilon\equiv t_s/L^3=1$ at the magnet-kink CQT, for several
  values of $L$.  We consider the Ising Hamiltonian at $h=0$ and $g =
  0.5$ with the bond perturbation (\ref{bondp}).  The curves are
  obtained by numerical calculations.\cite{footnotenum} They clearly
  approach an asymptotic function with increasing $L$, confirming
  OFSS.  The inset shows the same quantity at $\tau = 0$ versus
  $1/L$. }
\label{fchibond}
\end{figure}

We now study the off-equilibrium behavior arising when the system crosses
the CQT. We consider a time-dependent bond variable $b(t)$
such that $b(0) = b_c = 2$, obeying a linear time dependence:
\begin{equation}
\varepsilon(t) \equiv b(t) - 2 = -t/t_s. 
\label{varet}
\end{equation}
We assume that the evolution starts at time $t_i < 0$, so that
$\varepsilon(t_i)=\varepsilon_i>0$ in the kink phase.  Then, the
system evolves up to $t=t_f>0$ corresponding to
$\varepsilon(t_f)=\varepsilon_f<0$ in the magnet phase. Again we
expect an off-equilibrium behavior when $\varepsilon(t)$ changes sign,
which we describe using OS arguments analogous to those used in the
case of the single-site perturbation.

Using OS arguments analogous to those of the previous section, we
define the scaling variables
\begin{equation}
\varepsilon_t =- L\, t /t_s,\qquad \theta = t\,\Delta \sim t \,L^{-2},
\label{vartthetadefb}
\end{equation}
and also 
\begin{equation}
\upsilon = t_s L^{-3},\qquad \tau = t/t_s^{2/3}.  
\label{upstaudefb}
\end{equation}
In the limit $t,t_s,L\to \infty$ at fixed scaling variables, the
observables are expected to show OFSS. For example, we expect 
\begin{equation}
G[x_1,x_2;t,t_s,L] \approx m_0^2
\; {\cal G}(X_1,X_2;\upsilon,\tau),
\label{gscalo}
\end{equation}
and also 
\begin{equation}
\chi
\equiv \sum_x G(0,x;t,t_s,L) = m_0^2 \,L \,F_\chi(\upsilon,\tau).
\label{chiofss}
\end{equation}
Again, OFSS is confirmed by numerical computations,\cite{footnotenum}
and corrections appear to decay as $1/L$ (see Fig.~\ref{fchibond}).

\section{Summary and conclusions}
\label{conclu}

Summarizing, we have studied the effects of local time-dependent
perturbations of quantum many-body systems, focusing on phenomena
induced by large time scales $t_s$.  The off-equilibrium dynamics
close to quantum transitions obeys general scaling laws.  At a FOQT,
the behavior can be parametrized by the scaling variables $\upsilon =
\Delta^2 t_s$ and $\tau = t/\sqrt{t_s}$.  Some scaling functions can
be predicted, for large system sizes, using a two-level Hamiltonian
truncation.  For CQTs, analogous scaling variables can be defined,
that are uniquely specified by the standard critical exponents and by
the scaling dimension of the perturbation.  Moreover, at FOQTs local
variations of bond defects may lead to substantial changes of the bulk
low-energy properties, leading to a dynamic behavior which admits an
OS description, as well.  It is also possible to include the effect of
a small finite temperature, by adding the scaling variable $\rho =
T/\Delta$.

The OS framework depicted here has been explicitly worked out in the
quantum Ising model~\eqref{hedef}, but is quite general.  As a matter
of fact, it can be extended to any FOQT and CQT, providing information
on the possibility of controlling quantum phases, and their bulk
low-energy properties, by local changes.  Quite remarkably, the OFSS
behavior can be observed for relatively small sizes: in some cases a
limited number of spins already displays the asymptotic behavior (see,
e.g., Fig.~\ref{fsigmafig}).  Therefore, even systems of modest size
may show definite signatures of the OS scaling laws derived in this
work.  In this respect, present-day quantum-simulation platforms have
already demonstrated their capability to reproduce and control the
dynamics of quantum Ising-like chains with $\sim 10$ spins. Ultracold
atoms in optical lattices,\cite{Simon-etal-11} trapped
ions,\cite{Edwards-etal-10,Islam-etal-11,LMD-11,Kim-etal-11,Debnath-etal-16}
and Rydberg atoms\cite{Labuhn-etal-16} seem to be the most promising
candidates where the emerging universality properties of the quantum
many-body physics discussed here can be tested with a minimal number
of controllable objects.  Furthermore, in quantum computing, some
algorithms (notably the adiabatic ones) rely on a sufficiently large
gap,\cite{Bloch-08,YKS-10,AC-09} and thus fail at FOQTs.  The OS
theory that we presented may clarify how this occurs in finite
systems.

The OS arguments we developed can be extended to higher-dimensional
systems, such as 2D and 3D quantum Ising systems at their FOQTs and
CQTs, where novel features may arise depending on the various
possible geometries of the defects.
It is also tempting to generalize our framework to allow for dissipation,
such as that induced by the coupling with an external bath in a Markovian
framework.\cite{ABLZ-12,NVC-15,KMSFR-17} The emergence of novel 
intriguing scenarios may be tested in near-future experiments based 
on cavity-QED technology with superconducting
qubits.\cite{HBP-07,HTK-12,Fitz-etal-17}

\appendix

\section{Two-level reduction during the dynamics along the FOQT line}
\label{twoldyn}

Here we demonstrate that, similarly to EFSS, the dynamics in the OFSS limit
can be determined by using a two-level truncation of the Hamiltonian.
We consider a time-dependent Hamiltonian $H(t)$ and we assume that
\begin{equation}
   {\partial H \over \partial t} = {1\over t_s} A,
\end{equation}
where $A$ is independent of $t$. We recall that the
dynamics starts at $t = t_i < 0$ in one phase, and ends for $t = t_f >
0$ in the other phase. The transition point corresponds to $t=0$.  For
$t = t_i$ we require the system to be in the ground state of
Hamiltonian $H(t_i)$, which we can identify with $|-\rangle$ in the
large-volume limit.

To determine the dynamics, we should solve the evolution equation
\begin{equation}
 i {\partial \Psi\over \partial t} = H(t) \Psi.
\label{Sch}
\end{equation}
Let $\psi_n(t)$ and $E_n(t)$ be the orthonormalized eigenfunctions and
eigenvalues of $H(t)$. Here $\psi_0$ is the ground state of the system
and $\Delta_1(t) = E_1(t) - E_0(t)$.  We expand $\Psi(t)$ as
\begin{equation}
\Psi(t) = \sum_n c_n(t) \psi_n(t) e^{i\theta_n(t)}
\label{expansion}
\end{equation}
with 
\begin{equation}
\theta_n(t) = - \int_{t_i}^t E_n(s) ds .
\end{equation}
For $t=t_i$ we have $\Psi = \psi_0(t_i)$ and therefore 
$c_n(t_i) = \delta_{n0}$.
Substitution of the expansion (\ref{expansion}) into Eq.~(\ref{Sch}) gives 
\begin{equation}
{d c_n\over d t} = -
  \sum_k c_k 
  \left \langle \psi_n \left| {\partial \psi_k\over \partial t} \right. \right \rangle
  e^{i(\theta_k - \theta_n)},
\end{equation}
that must be solved with the boundary condition $c_n(t_i) = \delta_{n0}$. 
Differentiating the eigenvalue equation $H(t) \psi_n = E_n \psi_n$
with respect to $t$, we obtain
\begin{equation}
 \left \langle \psi_m 
  \left|{\partial H\over \partial t} \right| \psi_n\right\rangle = 
  (E_n - E_m) 
  \left \langle \psi_m \left| {\partial \psi_n\over \partial t}\right. 
   \right \rangle +
   \delta_{mn} {\partial E_n\over \partial t}.
\end{equation}
Therefore, we obtain
\begin{eqnarray}
{d c_n\over d t} &= &
  \sum_{k\not=n} 
   c_k {1\over t_s (E_n - E_k)} \langle \psi_n | A | \psi_k \rangle
  e^{i(\theta_k - \theta_n)} \nonumber \\
&- &
  c_n \left \langle \psi_n \left | {\partial \psi_n\over \partial t} \right.
  \right \rangle .
\label{eqcn}
\end{eqnarray}
If we just take the adiabatic limit $t_s\to \infty$, all cross terms 
can be neglected. Since $\psi_n$ is normalized, we can set 
\begin{equation}
\left \langle \psi_n \left| {\partial \psi_n\over \partial t}
    \right. \right \rangle = -i \phi_n(t)
\end{equation}
where $\phi_n(t)$ is a real function. Therefore, we have 
\begin{equation}
{d c_n\over d t} = i c_n \phi_n(t),
\end{equation}
whose solution, with the given boundary conditions, is simply 
$c_n(t) = 0$ for $n\ge 1$ and 
\begin{equation}
c_0(t) = \exp \left(i\int_{t_i}^t \phi_0(t) dt \right),
\end{equation}
which is nothing but the usual adiabatic theorem.\cite{Messiah}

In our case, however, the previous approximation does not work as we
are taking the limit at fixed 
\begin{equation}
t_s \, [E_1(t=0) - E_0(t=0)] \equiv t_s\,\Delta. 
\label{tsed}
\end{equation}
Thus, we must proceed more carefully. First, we note that the
differences $E_n(0)-E_0(0)$ and $E_n(0)-E_1(0)$ are strictly positive
in the FSS limit for any $n \ge 2$.  This implies that, in the FSS
limit, $dc_n/dt$ for $n\ge 2$ depends only on $c_k$ with $k\ge
2$. Given that all $c_n$ with $n\ge 2$ vanish for $t=t_i$, we can
conclude that we can set $c_n(t) = 0$ for all $n \ge 2$. On the other
hand, the coupling between the ground state and the first-excited state
cannot be neglected. Hence, in the OFSS limit the dynamics can be
determined by only considering two states, i.e., we can write
\begin{equation}
\Psi(t) = c_0(t) \psi_0(t) e^{i\theta_0(t)} + 
 c_1(t) \psi_1(t) e^{i\theta_1(t)}
\end{equation}
where $c_0(t)$ and $c_1(t)$ satisfy the coupled equations
\begin{eqnarray}
{d c_0\over d t} &=& 
   i c_0 \phi_0(t) 
    - {c_1 \over t_s \Delta_1(t)} \langle \psi_0|A|\psi_1\rangle 
    e^{i(\theta_1 - \theta_0)} , \;\\
{d c_1\over d t} &=& 
   i c_1 \phi_1(t) 
    + {c_0 \over t_s \Delta_1(t)} \langle \psi_1|A|\psi_0\rangle
    e^{-i(\theta_1 - \theta_0)}. \hspace*{0.5cm}
\end{eqnarray}
Corrections are of order $1/t_s$. Since $t_s \Delta$ is kept fixed in
the OFSS limit, corrections decrease as $\Delta$, that is
exponentially in the size of the system.

To make contact with the presentation in the paper, note that
$\psi_0(t)$ and $\psi_1(t)$ are the first two lowest states of the
model in the presence of a magnetic field. In the OFSS limit, as we
have stressed at the beginning, these two states can be written as
combinations of the magnetized states $|+\rangle$ and
$-\rangle$. Therefore, we can obtain the correct dynamic scaling
behavior by simply writing $\Psi(t) = e_0(t) |+\rangle + e_1(t)
|-\rangle$ and considering the evolution restricted to the subspace
spanned by these two states.  Corrections are again expected to be
exponentially small.


\begin{thebibliography}{99}
\bibitem{footnotenum} At the FOQTs, where a very high accuracy is
  needed in order to be able to discriminate exponential scalings, we
  calculated the magnetization by means of an exact diagonalization
  technique up to $L=7$.  At CQTs, we exploited the fact that the
  Hamiltonian only involves nearest-neighbor couplings, and simulated
  the dynamics of systems with up to 21 spins (the dimension of the
  Hilbert space is $\sim 2 \times 10^6$) by means of a Suzuki-Trotter
  decomposition of the unitary evolution operator as a product of
  local terms.  The time has been discretized in small steps of
  $\Delta t = 10^{-2}$.
\end{thebibliography}
\end{document}